\documentclass[letter]{aa} 

\makeatletter
\renewcommand*\aa@pageof{, page \thepage{} of \pageref*{LastPage}}
\makeatother
\usepackage{graphicx}
\usepackage{rotating}
\usepackage[varg]{txfonts}
\usepackage{txfonts}
\usepackage{color}
\usepackage{natbib}
\usepackage{amsmath}
\usepackage{hyperref}
\bibpunct{(}{)}{;}{a}{}{,} 

\def\bpic{$\beta$ Pictoris}

\def\cd{\,d$^{\rm -1}$}

\begin{document} 

   \title{Transiting exocomets detected in broadband light by TESS \\
   in the $\beta$ Pictoris system}
      \titlerunning{Transiting exoplanet around \bpic}

   \author{S. Zieba \inst{1}
         \and 
          K. Zwintz\inst{1}
          \and
          M. A. Kenworthy\inst{2}
          \and G. M. Kennedy \inst{3}
          }

   \institute{Universit\"at Innsbruck, Institut f\"ur Astro- und Teilchenphysik,
             Technikerstra{\ss}e 25, A-6020 Innsbruck\\
              \email{konstanze.zwintz@uibk.ac.at}
         \and
             Leiden Observatory, Leiden University, P.O. Box 9513, 2300 RA Leiden, The Netherlands
        \and
            Department of Physics, University of Warwick, Gibbet Hill Road, Coventry CV4 7AL, UK
             }

   \date{Received 26 March 2019; accepted }

  \abstract
   {}
   {We search for signs of falling evaporating bodies (FEBs, also known as exocomets) in photometric time series obtained for \bpic{} after fitting and removing its $\delta$ Scuti-type pulsation frequencies. }
   {Using photometric data obtained by the TESS satellite we determined the pulsational properties of the exoplanet host star \bpic{} through frequency analysis. We then pre-whitened the 54 identified $\delta$ Scuti $p$-modes and investigated the residual photometric time series for the presence of FEBs.}
   {We identify three distinct dipping events in the light curve of \bpic{} over a 105-day period. These dips have depths from 0.5 to 2 millimagnitudes and durations of up to 2 days for the largest dip. These dips are asymmetric in nature and are consistent with a model of an evaporating comet with an extended tail crossing the disc of the star.}
   {We present the first broadband detections of exocomets crossing the disc of $\beta$ Pictoris, complementing the predictions made 20 years earlier by \citet{lecavelier1999}. No periodic transits are seen in this time series. These observations confirm the spectroscopic detection of exocomets in calcium H and K lines that have been seen in high resolution spectroscopy.}


   \maketitle
%

\section{Introduction}

Our voyage of extrasolar planetary system discovery and exploration arguably began with the image of the circumstellar disc of $\beta$ Pictoris \citep{smith1984}, which showed that the infrared excesses discovered by the Infrared Astronomical Satellite  (IRAS) \citep[e.g.][]{1984ApJ...278L..23A} were not ``shells'', but discs on scales similar to the extent of our solar system.

$\beta$ Pictoris (HD\,39060) has a $V$ magnitude of 3.86 and lies at a distance of 19.76\,pc \citep{gaia-dr2}. It is a member of the $\beta$ Pictoris moving group with an age of $\sim$23\,Myr \citep{mamajek2014}. 
$\beta$ Pictoris harbours a warped debris disc composed of dust and gas, which is seen nearly edge-on and has an outer extent ranging from 1450 AU to 1835 AU \citep{larwood2001, apai2015}. In 2010, the planet $\beta$ Pictoris b was detected using observations with the NaCo instrument at the Very Large Telescope (VLT)  by \citet{lagrange2010}. 

Because of its close proximity and circumstellar disc, the $\beta$ Pictoris system can be considered an ideal test bed to study the formation and evolution of planetary systems, including minor bodies such as exocomets and exomoons. 
Indeed, the first reports of comets around stars other than our Sun came from spectroscopic observations of the $\beta$ Pictoris system, which identified transient absorption features that change in time \citep{ferlet1987,kiefer2014a}. Similar detections have also been made towards other stars, for example HD\,172555 \citep{kiefer2014b}. Models based on these detections have included cometary comae and have placed the exocomets at distances of a few to several tens of stellar radii, which was confirmed by their observed acceleration \citep{beust1990,kennedy2018}. Subsequent work theoretically predicted that occultations of the stellar light by these exocomets could be detected in broadband photometric time series. \citet{lecavelier1999} and \citet{lecavelier1999b} estimated that the depths of such signals would be of the order of tenths of a percent in flux and suggested the detection of an exocomet passage in front of $\beta$ Pictoris in observations from 1981 \citep{lecavelier1995, lecavelier1997, lamers1997}.

To date just a few stars show photometric transits consistent with the predictions. The first of these were discovered by \citet{Rappaport19} around the two stars KIC\,3542116 and KIC\,11084727. The dips seen for these stars show a remarkable similarity to the models of \citet{lecavelier1999}, which strengthens the exocomet interpretation. A more recent study by \citet{kennedy2019} describes an algorithm to identify transit features with an ingress that is more rapid than the egress, i.e. such as expected for exocometary bodies with a long tail. This work recovered the events discovered by \citet{Rappaport19}, found one more around HD\,182952, and argued that the host stars are likely young ($\sim$100Myr old). For completeness, we note that the deep and irregular dimming events seen towards KIC\,8462852 have also been interpreted as having a cometary origin \citep{boyajian2016,wyatt2018,kiefer2017}, although the shape and depth of these events are very different to those discovered by \citet{Rappaport19} and \citet{kennedy2018}.

In this letter we present observations of \bpic{} from TESS and show that the duration and depth of a dipping event in Sector 6 are consistent with the transit of an exocomet and agree with the predictions made by \citet{lecavelier1999}.

\begin{figure*}
\begin{center}
\includegraphics[width=0.9\textwidth]{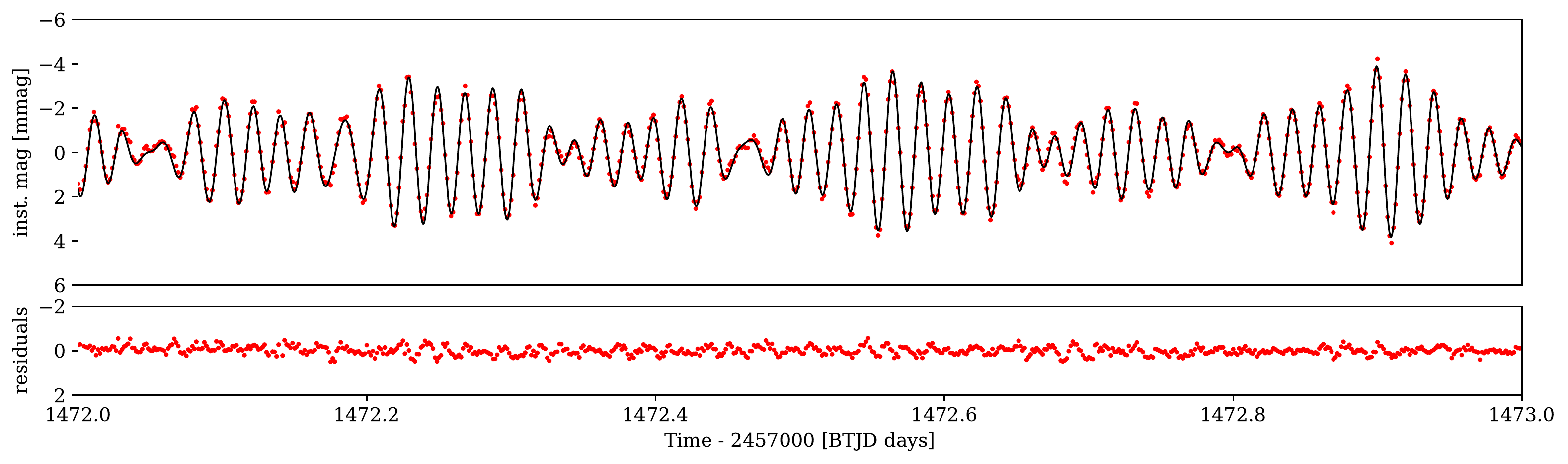}
\caption{One-day zoom of the \bpic{} light curve. {\it Upper panel:} TESS photometric time series (red points) and multi-sine fit using the 54 identified $\delta$ Scuti frequencies. {\it Lower panel:} Residual time series after subtracting the multi-sine fit using all 54 identified pulsation frequencies.}
\label{fig:puls_zoom}
\end{center}
\end{figure*}

\section{Observations}\label{sec:observations}

We used data collected by the Transiting Exoplanet Survey Satellite \citep[TESS;][]{ricker2015} from 19 October 2018 to 1 February 2019 in Sectors 4 through 7.
This leads to a baseline of about 105 days with a duty cycle of 85\% for the four sectors.
\bpic{} (TIC\,270577175, T = 3.82 mag) is one of the preselected targets for which ``short cadence'' 2-minute data are provided.
We investigated the instrumental behaviour of TESS during the $\beta$ Pictoris observations in detail to exclude any systematic effects or instrumental artefacts. A detailed description is given in the Appendix.

\section{Results}

\subsection{Fitting the stellar pulsations of Beta Pictoris} \label{sec:pulse}

\bpic{} is known to show $\delta$ Scuti-like pulsations \citep{koen03a,koen03b,mekarnia17,zwintz19}, which we needed to fit and remove in order to search for signs of exocomets. 
Therefore, we conducted a frequency analysis using the software package {\tt Period04} \citep{lenz05} and extracted all pulsation frequencies down to a signal-to-noise ratio of 4 following \citet{breger1993}.
From this, we identified 54 significant $p$-modes in the frequency range from 23 \cd  and 76\cd. A one-day zoom into the light curve illustrating the pulsational behaviour is shown in the upper panel of Figure \ref{fig:puls_zoom}.
A detailed description, the full list of frequencies (see Table \ref{tbl:pulsations}) and the corresponding amplitude spectrum (see Figure \ref{fig:ampspec}) are provided in the Appendix.
We subtracted the multi-sine fit using the 54 frequencies, amplitudes, and phases from the original TESS light curve and investigated the residuals (lower panel in Figure \ref{fig:puls_zoom}) for the presence of FEBs.

\begin{figure*}
\begin{center}
\includegraphics[width=0.95\textwidth]{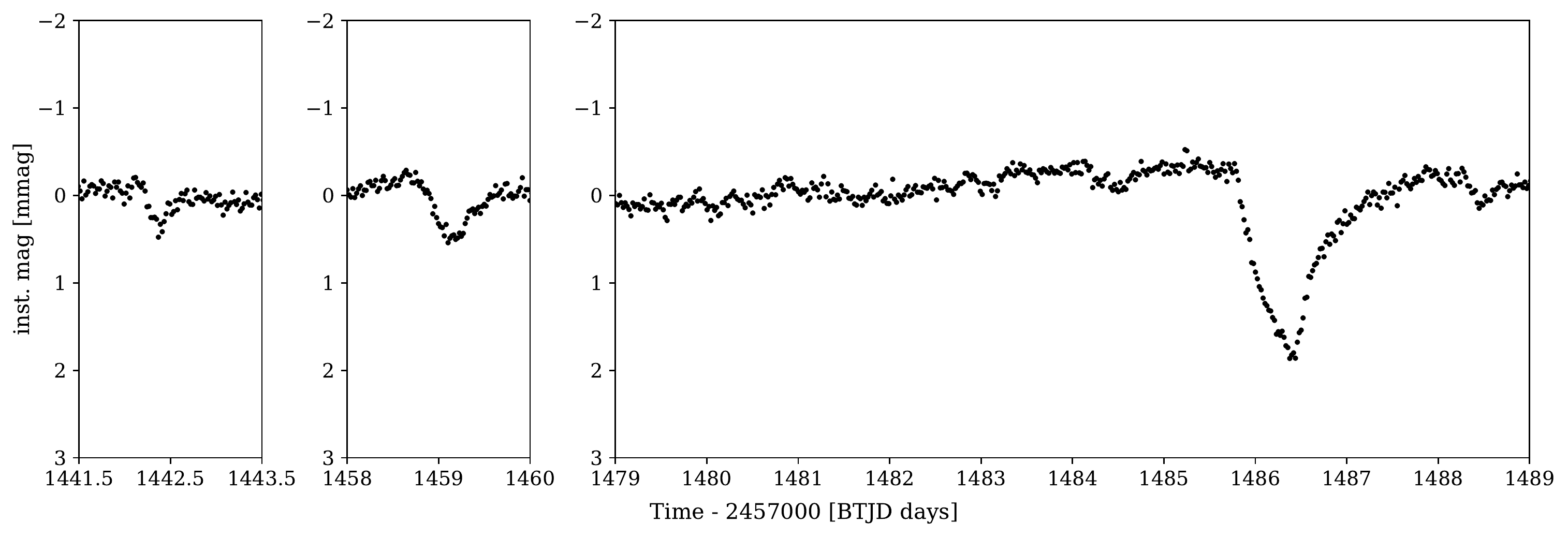}
\caption{Zoom-in of the three events (from left to right: transit C, B and A) in the 30-minute binned light curve after subtraction of the pulsational signal.}
\label{fig:3zoom}
\end{center}
\end{figure*}

\subsection{Exocomet transit modelling}\label{sec:transit}

The pulsation removed light curve shows three notable transit features, as highlighted in Figure~\ref{fig:3zoom}.
All three transits show a triangular or sawtooth shape that has a steep increase of absorption followed by an exponential decay back to the full flux level of the star.
The three transits have different shapes and are not periodic. Assuming a circular orbit with period $P>105$ days and a mass of $1.80 M_\odot$  for $\beta$ Pictoris \citep{Wang16} leads to a minimum orbital radius of about 0.5 AU. It is almost certain that the comet orbits are highly eccentric, so this non-periodicity is not particularly surprising.

\citet{lecavelier1999} carried out a comprehensive modelling study of exocomet light curves that includes modelling of the tail ejecta and its particle size distribution and the subsequent acceleration due to radiation pressure from the star for different distributions of the sizes of the ejected particles.
Not knowing the geometry of the exocomet orbit leads to degeneracies in the possible range of particle distributions and orbital parameters, so in this Letter we adopt a simpler 1-D model as used in \citet{Brogi12}. While that model was designed with a transiting planet in mind, the only practical difference in this work is that the orbital period and eccentricity are largely unconstrained. We therefore assume a circular orbit with period $P$ and orbital radius $a$ for the modelling, but report the results in terms of the transverse velocity $v_{trans}$ of the comet across the face of the star.
The angle between the observer, centre of the star, and the comet $\theta$ is related to the time $t$ by $\theta= 2\pi(t-t_{mid})/P$, where $t_{mid}$ is the time of the transit following the nomenclature of \citet{Brogi12}. The star is modelled as a limb-darkened circular disc with radius $R_*=1.53 R_\odot$ \citep{Wang16} and linear limb darkening coefficient $u=0.275$ for $\beta$ Pictoris \citep{Claret00}.

The exocomet is assumed to be optically thin, and its extinction cross-section $\rho$ is expressed in units of stellar area. 
The model has a hard front edge and the cross section then drops exponentially  behind the comet with normalisation constant $c_e$ and an exponential factor $\lambda$ so that $\rho(\Delta \theta)=c_e e^{-\lambda(\Delta \theta)}$, where $\Delta \theta=(\theta - \theta')$ is the angle between the position of the comet and an arbitrary point on the circular orbit of the comet.
The characteristic length of the tail is therefore $1/\lambda$.

The comet crosses the limb-darkened disc of the star with an impact parameter $b$ and chord length $r_c$, related by $b=[1-(r_c/2R_*)]^{1/2}$.
The total extinction of the star is the convolution of $\rho$ with the stellar disc, and so the resultant intensity $I$ of the star is

\begin{equation}
I(\theta) = 1 - \int^{2\pi}_0 \rho(\theta-\theta') i (\theta',\hat{r}_c)\,d\theta'.
\end{equation}

The function $i(\theta',\hat{r}_c)$ is the limb-darkened intensity of the star along a given chord defined by
\begin{equation}
i(\theta',\hat{r}_c)=1-u\left [ 1- \frac{a}{R_*}\sqrt{\sin^2(\hat{r}_c/2)-\sin^2 \theta'} \right ],
\end{equation}

which is set to zero for $\hat{r}_c/2<\theta ' < 2\pi - \hat{r}_c/2$, where
$\theta'$ is the angle between the centre of the star and a point along the orbit of the comet and the angular size of the chord as seen from the comet $\hat{r}_c= \arcsin(r_c/2a)$.

 With five parameters ($t_{mid}$, $b$, $c_e$, $\lambda$, $P$)  the model is underconstrained in that transits with more rapid ingress at large $b$ (shorter transit chords) are compensated for by longer orbital periods and shorter comet tails, thus yielding the same ingress or egress slope in the model light curve. Fixing $b=0$ constrains $P$ and $\lambda$, although if the comet were not assumed to have a hard front edge these would again be degenerate.

\begin{figure}
\begin{center}
\includegraphics[width=0.40\textwidth]{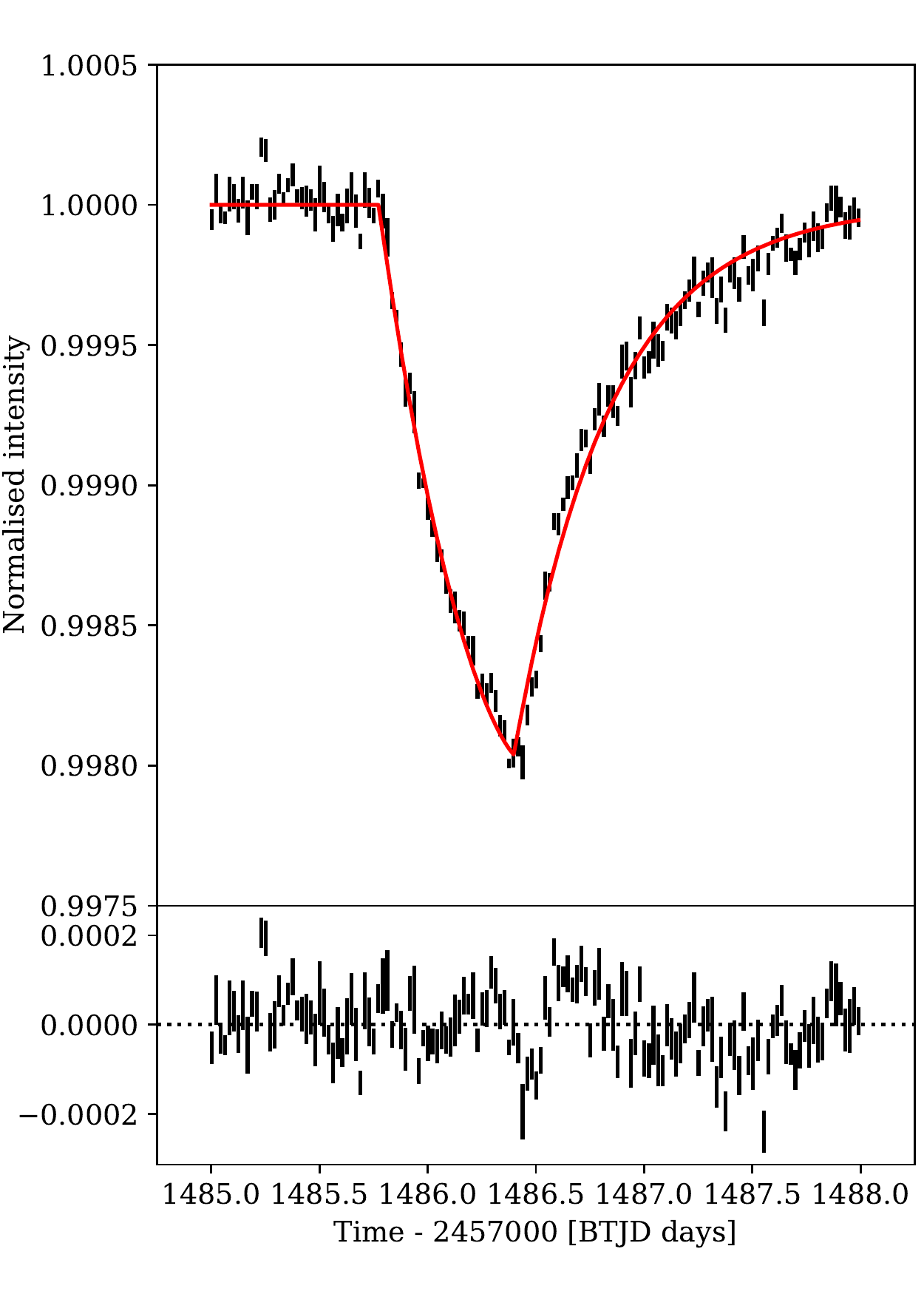}
\caption{Best-fit comet model. Upper panel: The binned photometry showing the largest transit event. The vertical dashes show the error bars on the photometry. The red line shows the best-fit model for $b=0$ (i.e. the median of the parameters shown in Fig. \ref{fig:triang}). The lower panel shows the residuals from the fit.}
\label{fig:dip_and_fit}
\end{center}
\end{figure}

\begin{figure}
\begin{center}
\includegraphics[width=0.45\textwidth]{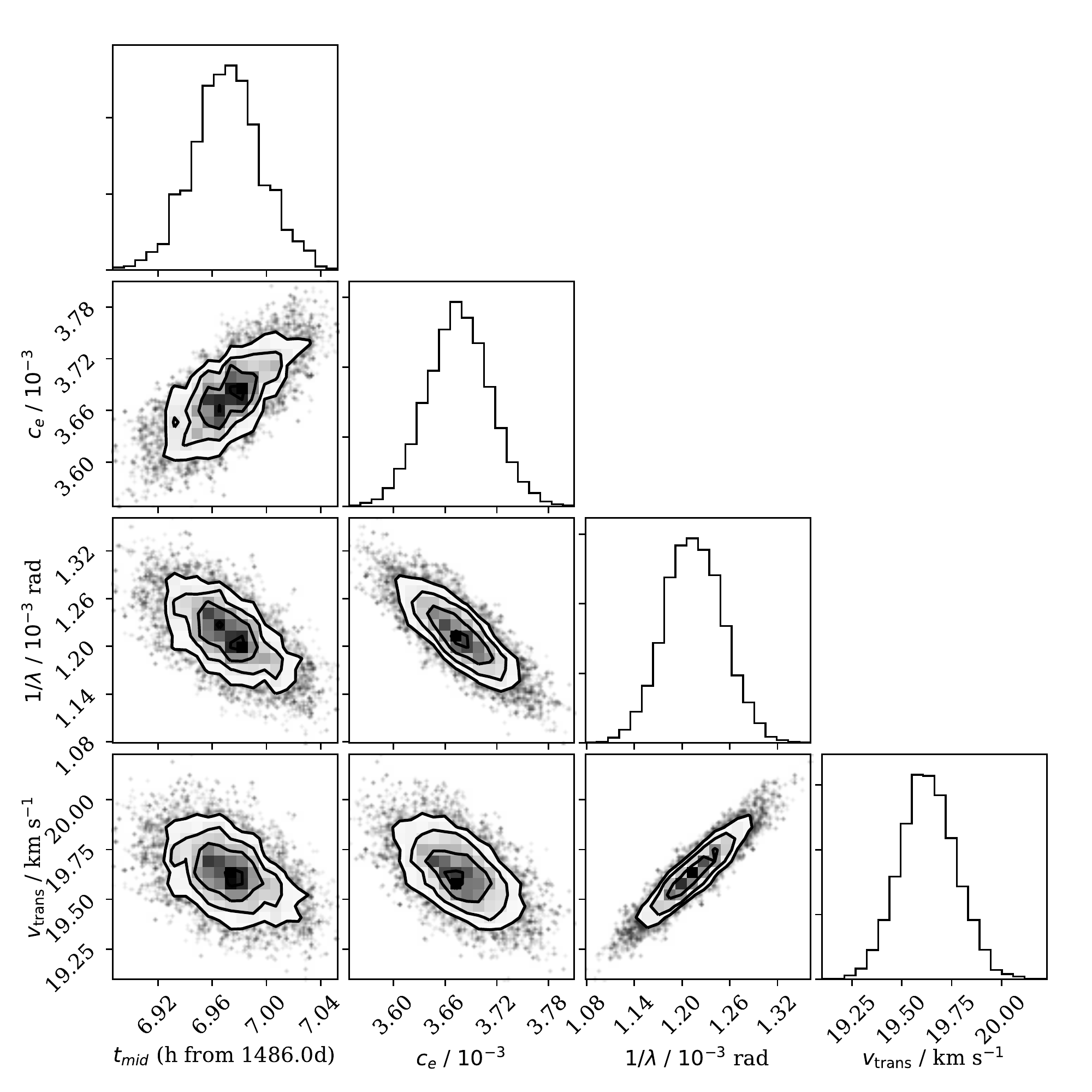}
\caption{Posterior probability distributions for the exocomet model with $b=0$. Diagonal panels show the 1-D distributions, and the other panels show the 2-D distributions and illustrate the parameter degeneracies (primarily between transverse velocity and comet tail length).}
\label{fig:triang}
\end{center}
\end{figure}

We use the Python software package {\tt emcee} \citep{Mackey13} to perform a Markov Chain Monte Carlo (MCMC) fit to the binned light curve around the largest transit feature.
We average the photometry over 30-minute bins, and calculate the error bars on the binned photometry by determining the root mean square of each set of photometric points and dividing by the square root of the number of points per bin.
The photometric light curve and the best-fit model are shown for $b=0$ in Figure~\ref{fig:dip_and_fit}, along with the residuals of the fit in the lower panel. The resultant triangle plot showing the parameter distributions and correlations between the variables is shown in Figure~\ref{fig:triang}.
The $\pm1\sigma$ errors of the parameters are computed by measuring the interval between 16\% and 84\% of the merged distribution, and the best-fit values and their uncertainties are listed in Table~\ref{tbl:params}.

\begin{table}
\caption{Fitted parameters from the MCMC routine with $b=0$.}
\label{tbl:params}
\small
\def\arraystretch{1.2}
\setlength{\tabcolsep}{8pt}
\centering
\begin{tabular}{@{}lll@{}}
\hline\hline
Parameter & Value \\
\hline
$t_{mid}$  & $1486.290 \pm 0.001$ d  \\
$b$        & $0$ (assumed)  \\
$c_e$      & $(3.67 \pm 0.04) \times 10^{-3}$ \\
$1/\lambda$ & $(1.22 \pm 0.04) \times 10^{-3}$ rad  \\
$v_{trans}$   &  $19.6 \pm 0.1$ km\,s$^{-1}$   \\
\hline\hline
\end{tabular}
\end{table}

\section{Discussion}

As noted above, the comet orbit is not constrained by our model because the chord across the star is unknown, yielding a near-total degeneracy between $b$, $\lambda$ and $v_{trans}$, which is derived from $P$. By setting $b=0$ the transverse velocity is constrained, for our assumption of a hard front edge for the comet, and has a value of $v_{trans}=19.6{\rm~km~s}^{-1}$ (i.e. equivalent to the circular velocity at about 3 AU). This transverse velocity is surprisingly slow given that the radial velocities of the comets detected in calcium absorption are typically tens to hundreds of km\,s$^{-1}$ and imply distances at transit of only a few tens of stellar radii. This discrepancy therefore implies a high eccentricity (i.e. near-radially plunging) orbit if the spectroscopic and photometric events were from objects on similar orbits. While high eccentricity orbits are expected, a further issue is that the characteristic transit shape seen in this work and by \citet{Rappaport19} is not expected for comets with arbitrary pericentre orientations \citep{lecavelier1999,lecavelier1999b}. Further orbital constraints may therefore be derived from the TESS data, which we leave for more detailed future modelling work. 

The characteristic tail length $1/\lambda$ is small. However, as with the velocity this is also a transverse length, and in addition the physical size depends on the distance to the star. If we assume 1~AU, then this length is $\sim$$2 \times 10^8$~km, or about a tenth of the stellar diameter, indicating that most of the dust is concentrated near the comet nucleus. The close match of the data with the models of \citet{lecavelier1999} suggests that those physical models are a reasonable starting point to further interpret these data.

We chose not to include forward scattering for our exocomet model, even though \citet{lecavelier1999} and others include it in their models.
Our hesitation is that the long-term variation in the flux of \bpic{} over Sector 6 seems to not extend or be seen in the adjacent sectors.
An initial examination suggests that there is a long-lived forward scattering halo associated with transit A (illustrated in the rightmost panel of Figure \ref{fig:3zoom}), but it is not centred on the transit itself.
Forward scattering peaks are seen in transit B and C. However because these transits are at much smaller significance than transit A, there is a much larger degeneracy in the model fitting for these two dips, and exploring whether these dips have a common orbital connection with transit A is left to a future paper.

\section{Conclusions}

We report the first photometric broadband detection of an exocomet transit in the \bpic{} system, consistent with the prediction of \citet{lecavelier1999}.
We fit a simple cometary tail model to the largest of these events, seen at BTJD 1486.4, and find that it is consistent with an exponentially decaying optical depth tail convolved with the limb-darkened disc of the star.

Future work includes a more comprehensive modelling of all the transits seen in the TESS light curve, and a  study of whether high resolution spectral observations were taken simultaneously with this transit to confirm the cometary hypothesis.
A comprehensive photometric campaign was conducted during the Hill sphere transit of $\beta$ Pictoris b \citep{Kenworthy17}, which includes bRing \citep{Stuik17}, ASTEP   \citep{Guillot15,Abe13,mekarnia17}, and BRITE-Constellation \citep{weiss2014} photometry.
Searching this data may yield more candidate broadband transits.

A dedicated cubesat mission to monitor \bpic{}, similar to the PICSAT project  \citep{Nowak16,Nowak18} would uncover more of these events, and photometry at different wavebands would help determine the size distribution of the tail particles \citep{lecavelier1999}.

\begin{acknowledgements}

This paper includes data collected by the TESS mission, which are publicly available from the Mikulski Archive for Space Telescopes (MAST).
Funding for the TESS mission is provided by the NASA Explorer Program.
This work has made use of data from the European Space Agency (ESA) mission {\it Gaia} (\url{https://www.cosmos.esa.int/gaia}), processed by the {\it Gaia} Data Processing and Analysis Consortium (DPAC; \url{https://www.cosmos.esa.int/web/gaia/dpac/consortium}).
Funding for the DPAC has been provided by national institutions, in particular the institutions participating in the {\it Gaia} Multilateral Agreement.
GMK is supported by the Royal Society as a Royal Society University Research Fellow.
We made use of the software package {\tt Period04} \citep{lenz05}, the {\tt Python} programming language \citep{rossum1995}, and the open-source {\tt Python} packages {\tt numpy} \citep{walt2011}, {\tt matplotlib} \citep{hunter2007}, {\tt astropy} \citep{astropy2013}, {\tt lightkurve} \citep{barentsen2019}, and {\tt emcee} \citep{Mackey13}.
An on-line repository with materials used in this work is available at \url{https://github.com/sebastian-zieba/betaPic_comet}.

\end{acknowledgements}

\bibliographystyle{aa}
\bibliography{exocomet_lang.bib}

\appendix

\section{Instrumental behaviour}
\label{sec:inst}

\subsection{TESS photometric time series}
\bpic{} (TIC\,270577175, T = 3.82 mag) is one of the preselected targets for which ``short cadence'' 2-minute data is provided. This Candidate Target List (CTL) is a subset of the TESS Input Catalogue\footnote{\url{https://mast.stsci.edu/portal/Mashup/Clients/Mast/Portal.html}} \citep[TIC;][]{stassun2018} with about 200\,000 targets for TESS 2-minute cadence observations, which were primarily chosen in order to maximise the yield of transiting exoplanets.

\begin{figure*}
\begin{center}
\includegraphics[width=0.9\textwidth]{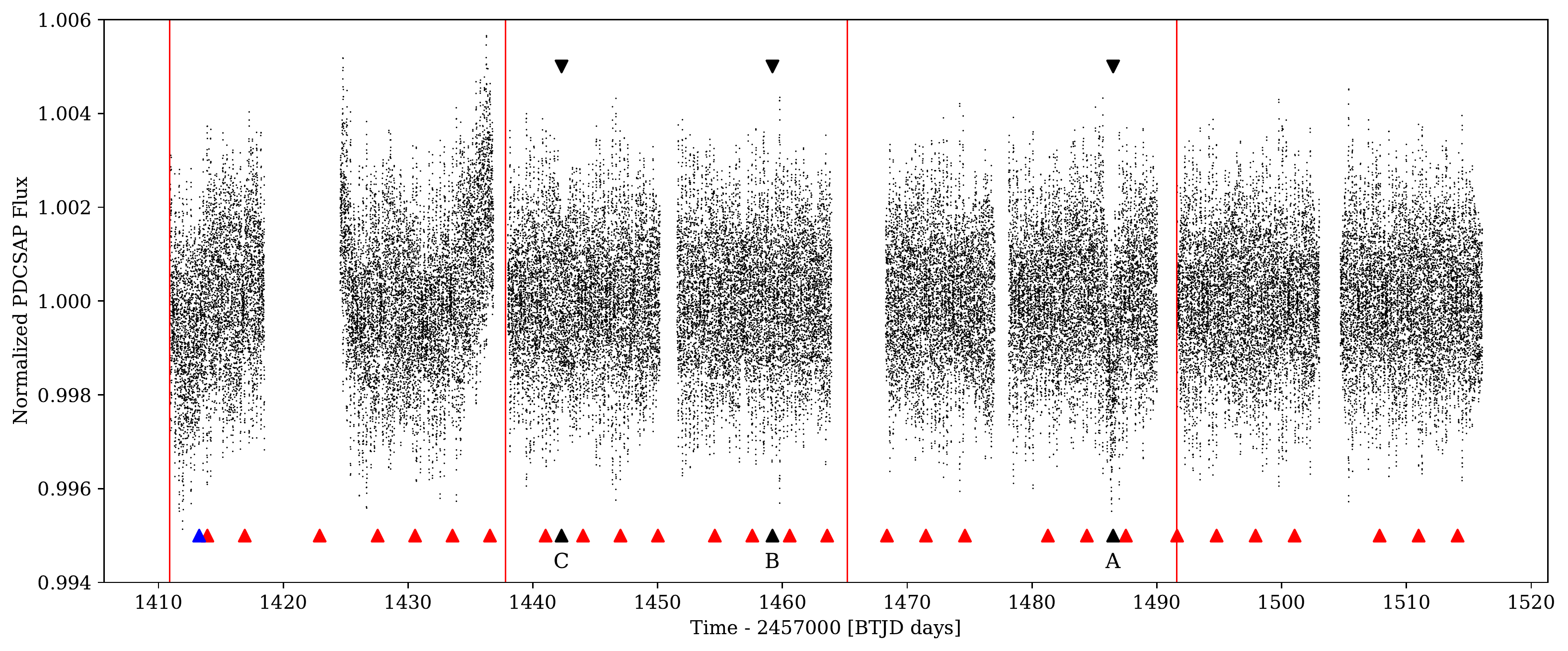}
\caption{
Full PDCSAP light curve of \bpic{} used in this analysis. The red vertical lines indicate the starts of Sectors 4 through 7. The black
triangles point to the three identified decreases in flux, labelled A,B,C in decreasing signal to noise. The time of momentum dumps are indicated
with red triangles; a small change in spacecraft pointing at the beginning of Sector 4 (due to an updated guide star table) is shown with a blue triangle.
}
\label{fig:full_lc}
\end{center}
\end{figure*}

Figure \ref{fig:full_lc} shows the complete TESS \bpic{} light curve.
The beginnings of each  sector are delineated with a vertical red line.
The visible gaps are related to the data downlink at the perigee of the orbit of TESS, which occurs every 13.7 days and during which TESS halts its observations for about one day.
\bpic{} was observed with CCD 1 of Camera 4 during Sector 4, with CCD 4 of Camera 3 during Sector 5 and with CCD 3 of Camera 3 during Sectors 6 and 7.
As the angle between the boresight of these cameras and both the Earth and the Moon was never smaller than $37^{\circ}$, we do not expect strong features created by scattered light from these two bodies during these observations. 

We used the 2-minute Presearch Data Conditioning \citep[PDC;][]{smith2012,stumpe2012} light curve from the Science Processing Operations Center (SPOC) pipeline \citep{jenkins2016,jenkins2017}, which was originally developed for the \textit{Kepler} mission \citep{jenkins2010}.
These light curves were corrected for systematics by the SPOC pipeline.
During the ``momentum dumps'', i.e. thruster firings that reduce the speed of the reaction wheels and decrease the pointing jitter, TESS loses the fine attitude control mode for about 15 minutes (see the TESS Instrument Handbook\footnote{\url{https://archive.stsci.edu/missions/tess/doc/TESS_Instrument_Handbook_v0.1.pdf}}).
Consequently, the pointing is less stable during these particular times, resulting in potential changes in the photometric fluxes of the target stars.
Momentum dumps took place every $\approx$2.5 days in Sector 4, every three days in Sector 5, and every 3.125 days in Sector 6 and 7 (red triangles in Figure \ref{fig:full_lc}).
We omitted data collected in the vicinity of momentum dumps in the subsequent analysis.

The data products were accessed and modified with the python package {\tt lightkurve} \citep{barentsen2019}, which retrieves the data from the MAST archive\footnote{\url{https://archive.stsci.edu/tess/}}.
We removed every measurement with a non-zero ``quality'' flag (see §9 in the TESS Science Data Products Description Document\footnote{\url{https://archive.stsci.edu/missions/tess/doc/EXP-TESS-ARC-ICD-TM-0014.pdf}}), which marks anomalies like cosmic ray events or instrumental issues. 
Three days between BTJD 1421 and BTJD 1424 were removed owing to higher than normal rates of spacecraft jitter.
This occurred just after an instrument anomaly (see the Data Release Note of Sector 4 for more information\footnote{\url{https://archive.stsci.edu/tess/tess_drn.html}}) and is visible in Figure \ref{fig:full_lc} as the first gap, which is noticeably bigger than the others. 
Finally, we normalised the four sectors by dividing each of the sectors by their respective median flux and we combined these into one light curve.

\subsection{Identification of possible transit events}

\begin{figure}
\begin{center}
\includegraphics[width=0.49\textwidth]{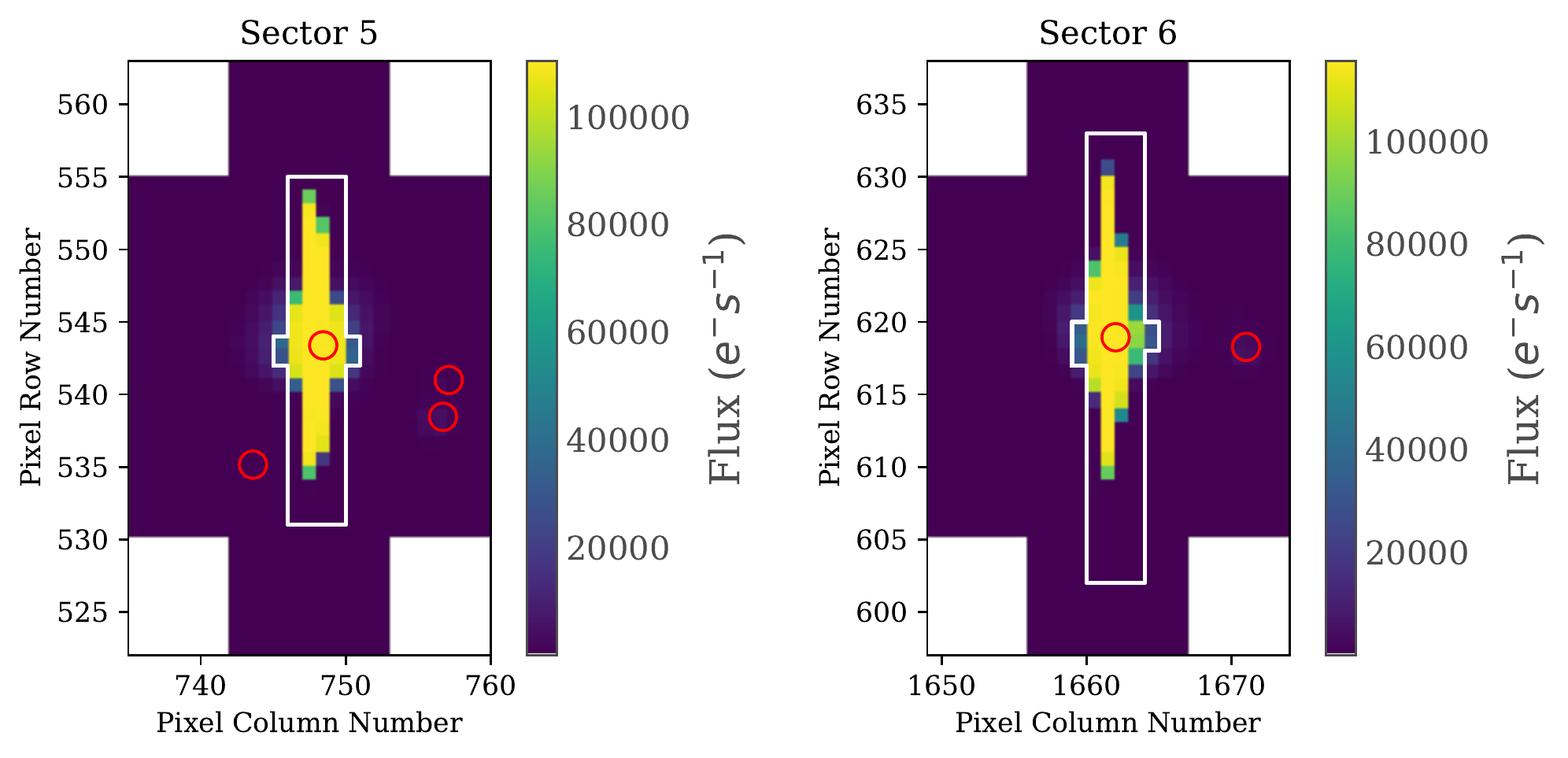}
\caption{Postage stamps of \bpic{} for Sector 5 and 6.
The stars which are bright enough to create such a dip in Sector 5 and 6, respectively, are delineated with red circles.
The white box indicates the pipeline aperture for those sectors.
The only star in the aperture is in both cases \bpic{}.}
\label{fig:betaPic_otherstars}
\end{center}
\end{figure}

Sector 5 saw an automated creation of a data validation (DV) report, where the pipeline identified two decreases in brightness at about BTJD 1442 and BTJD 1459.
These dips are indicated by the leftmost two downward-pointing black arrows in Figure \ref{fig:full_lc}.
The third arrow represents a clearly visible change in brightness for about two days at BTJD 1486.
This arrow is asymmetric in shape and very similar to the predicted light curve generated by a comet with an extended tail transiting the disc of the star, as hypothesised by \citet{lecavelier1999} and \citet{lecavelier1999b}.
We used the apertures which were created by the pipeline visible as the white boxes in Figure \ref{fig:betaPic_otherstars}.
The apertures have an elongated shape due to the blooming effect with a typical size of about 30 rows by 4 columns.
To determine the robustness of these measurements, custom apertures of differing geometries were used to extract the photometry of \bpic{} - all three dips appear in these reductions.
As a quick check, we used the TESScut function of {\tt lightkurve}, which allows us to create a light curve out of the full frame images (FFIs).
Those products are generated every 30 minutes by TESS.
However, we have to consider that those - in contrast to the light curves provided by the pipeline - do not have an aperture mask and are not background corrected.
After trying out different aperture masks and background masks for \bpic{}, we report that the dimming features are also distinctly visible in the FFIs. 

We then inspected the postage stamp images, which have a cadence of 2 minutes.
As the background is subtracted from the flux of the star, a bright object passing by, such as an asteroid,  could temporarily increase the background flux and thus decrease the flux of the star.
However, nothing like this is visible in the postage stamps during the dimming events.
The times of the event do not coincide with momentum dumps (as can be seen in Figure \ref{fig:full_lc}) and the jitter of TESS also did not increase.

We checked for similar dimmings in the stars closest to \bpic{} that have short cadence data.
TIC 270574544 and TIC 270626557 have a separation of 810 and 1000 arcseconds relative to \bpic{}, which corresponds to 39 and 48 pixels, respectively.
Those two stars were observed in the same sectors as \bpic{} and on the same CCDs and no similar decrease in brightness is visible.
Different background apertures in the vicinity of \bpic{} show no dips either.
These last two checks also rule out a ``rolling shutter'' effect.

The brightest star besides \bpic{} in the pipeline aperture is Gaia DR2 4792773113918130560 in Sector 5 and Gaia DR2 4792772873399766528 in Sector 6 \citep{gaia2016}. 
Those stars are too faint to create those dimming events. 
Stars in the postage stamps of \bpic{} with a luminosity high enough to create those dips are indicated with red circles in Figure \ref{fig:betaPic_otherstars}.

Finally we checked the possibility of Intra Camera Crosstalk as the cause of the dimmings. 
A bright object can create a decrease in flux in other parts of the same CCD (see §6.8.6 of the TESS Instrument Handbook\footnote{\url{https://archive.stsci.edu/missions/tess/doc/TESS_Instrument_Handbook_v0.1.pdf}}).
Those echoes can be found on the same row and 1024 columns higher than the bright source and they were only seen in CCD 3 of each camera.
The big dip in Sector 6 was observed with CCD 3, however 1024 columns above \bpic{} lies outside the imaging area of this CCD.
We conclude that \bpic{} is the true host of the dimming events.

\section{Details on the pulsation frequency analysis}
For the pulsation frequency analysis we removed all Sector 4 data as the PDCSAP flux exhibits strong variations.
As these are not visible in the SAP light curve, they are most likely artefacts of the reduction.
Consequently, this part of the light curve would introduce artificial frequencies and, hence, was omitted for the pulsational analysis.
The total time baseline of the TESS light curve used for the pulsational analysis is 78 days with a duty cycle of 80.5\%.

\begin{figure*}
\begin{center}
\includegraphics[width=0.9\textwidth]{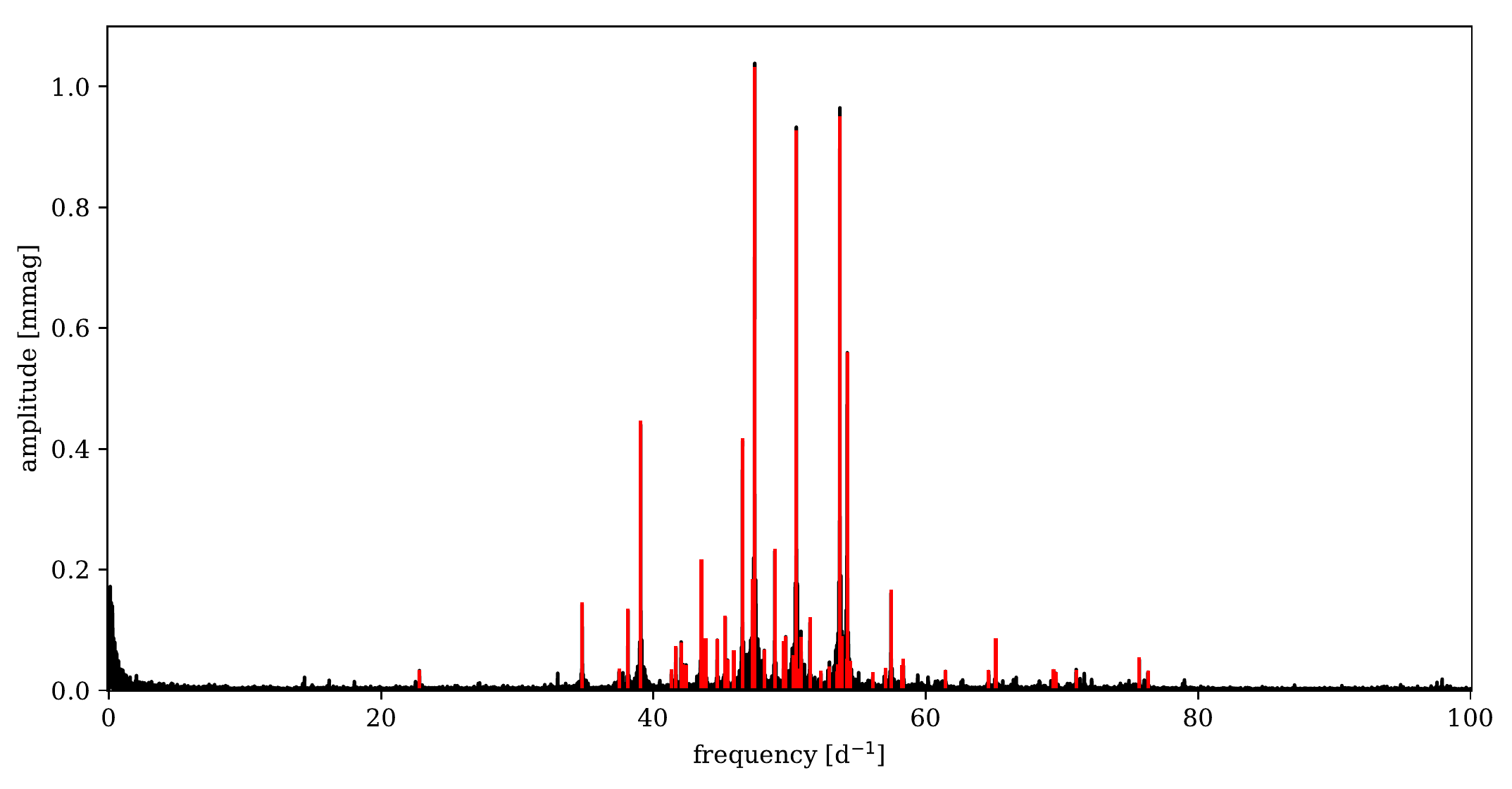}
\caption{Pulsation frequency analysis of \bpic{}. The original amplitude spectrum is shown in black and the 54 identified $\delta$ Scuti pulsations are shown in red.}
\label{fig:ampspec}
\end{center}
\end{figure*}

The original amplitude spectrum of $\beta$~Pictoris based on TESS data is shown in Figure \ref{fig:ampspec}, where the detected pulsation frequencies are represented in red.
The 54 $\delta$\,Scuti-type frequencies, amplitudes and phases detected in the TESS data for $\beta$~Pictoris are listed in Table \ref{tbl:pulsations}. Frequency, amplitude and phase errors are calculated using the relations by \citet{Montgomery1999}.
The zero point for the phase calculation is at BTJD = 1410.9035\,d.

\begin{table}
\caption{Pulsational frequencies, amplitudes, phases, and signal-to-noise values sorted by the pre-whitening sequence. Last-digit errors in frequencies, amplitudes, and phases are given in brackets.}
\label{tbl:pulsations}
\small
\def\arraystretch{1.2}
\setlength{\tabcolsep}{8pt}
\centering
\begin{tabular}{@{}lllll@{}}
\hline\hline
\# & Frequency        & Amplitude & Phase & S/N \\
 & $[{\rm d}^{-1}]$ & [mmag]    &       &     \\
\hline
1 & 47.438928(12) &     1.048(2) &      0.9334(4) &     34.3 \\
2 & 53.691744(12) &     1.048(2) &      0.1797(4) &     36.4 \\
3 & 50.491833(13) &     0.940(2) &      0.3251(4) &     39.6 \\
4 & 54.23744(2) &       0.568(2) &      0.5686(7) &     41.7 \\
5 & 39.06304(3) &       0.442(2) &      0.8625(9) &     40.4 \\
6 & 46.54302(3) &       0.414(2) &      0.7582(9) &     27.3 \\
7 & 48.91878(5) &       0.230(2) &      0.5943(17) & 19.1 \\
8 & 43.52779(6) &       0.217(2) &      0.7506(18) & 20.6 \\
9 & 47.28386(7) &       0.175(2) &      0.284(2) & 22.1 \\
10 & 57.45209(8) &      0.163(2) &      0.203(2) & 27.0 \\
11 & 34.76041(9) &      0.143(2) &      0.914(3) & 37.4 \\
12 & 38.12911(9) &      0.133(2) &      0.789(3) & 33.2 \\
13 & 45.26957(10) &     0.122(2) &      0.705(3) & 15.9 \\
14 & 51.49625(11) &     0.118(2) &      0.327(3) & 15.5 \\
15 & 47.27019(12) &     0.108(2) &      0.111(4) &      15.5 \\
16 & 53.85463(14) &     0.090(2) &      0.419(4) &      13.0 \\
17 & 49.71250(16) &     0.077(2) &      0.218(5) &      12.9 \\
18 & 50.83102(15) &     0.087(2) &      0.612(4) &      15.7 \\
19 & 43.82885(15) &     0.083(2) &      0.068(5) &      12.9 \\
20 & 65.13492(15) &     0.082(2) &      0.288(5) &      24.1 \\
21 & 44.68340(15) &     0.083(2) &      0.151(5) &      16.9 \\
22 & 49.55926(16) &     0.079(2) &      0.297(5) &      14.9 \\
23 & 42.03524(16) &     0.077(2) &      0.173(5) &      16.7 \\
24 & 41.65028(17) &     0.073(2) &      0.850(5) &      18.2 \\
25 & 48.1378(2) &       0.062(2) &      0.744(6) &      15.4 \\
26 & 45.90034(20) &     0.064(2) &      0.896(6) &      14.3 \\
27 & 50.2686(2) &       0.056(2) &      0.228(7) &      15.8 \\
28 & 75.6781(2) &       0.051(2) &      0.416(8) &      17.2 \\
29 & 58.3472(3) &       0.041(2) &      0.117(9) &      9.2 \\
30 & 54.2283(2) &       0.062(2) &      0.411(6) &      9.2 \\
31 & 45.4369(3) &       0.049(2) &      0.910(8) &      13.4 \\
32 & 54.4625(3) &       0.048(2) &      0.198(8) &      9.5 \\
33 & 53.5523(3) &       0.038(2) &      0.143(10) &     12.6 \\
34 & 42.1729(3) &       0.044(2) &      0.557(9) &      12.5 \\
35 & 58.2512(3) &       0.041(2) &      0.309(9) &      13.7 \\
36 & 42.3960(3) &       0.040(2) &      0.16(1) &       4.7 \\
37 & 52.9223(3) &       0.040(2) &      0.54(1) &       9.0 \\
38 & 53.68985(7) &      0.174(2) &      0.607(2) &      15.5 \\
39 & 57.0484(4) &       0.032(2) &      0.928(12) &     15.4 \\
40 & 50.6455(4) &       0.033(2) &      0.050(12) &     6.7 \\
41 & 37.4786(4) &       0.033(2) &      0.636(11) &     6.6 \\
42 & 69.3752(4) &       0.033(2) &      0.357(12) &     13.7 \\
43 & 41.3189(4) &       0.033(2) &      0.422(12) &     18.9 \\
44 & 61.4461(4) &       0.032(2) &      0.434(12) &     4.3 \\
45 & 22.8145(4) &       0.031(2) &      0.946(12) &     4.3 \\
46 & 64.6154(4) &       0.030(2) &      0.763(13) &     4.2 \\
47 & 71.0441(4) &       0.036(2) &      0.016(11) &     4.2 \\
48 & 52.3034(4) &       0.031(2) &      0.909(12) &     8.8 \\
49 & 47.4101(4) &       0.035(2) &      0.914(11) &     10.0 \\
50 & 76.3167(4) &       0.030(2) &      0.898(13) &     10.4 \\
51 & 45.3553(4) &       0.028(2) &      0.307(14) &     10.1 \\
52 & 53.4991(4) &       0.031(2) &      0.764(12) &     8.2 \\
53 & 69.5526(5) &       0.027(2) &      0.687(14) &     4.6 \\
54 & 56.1099(5) &       0.026(2) &      0.005(15) &     4.0 \\
\hline
\end{tabular}
\end{table}

\end{document}